\documentclass[prb,10pt,twocolumn,superscriptaddress]{revtex4}

\usepackage{epsfig}
\graphicspath{{/}}
\usepackage{graphicx}
\usepackage{slashed}
\usepackage{mathtools}

\begin{document}

\title{Exciton-Trion-Polaritons in Doped Two-Dimensional Materials}

\author{Farhan Rana}
\address{School of Electrical and Computer Engineering, Cornell University, Ithaca, NY 14853}
\author{Okan Koksal}
\address{School of Electrical and Computer Engineering, Cornell University, Ithaca, NY 14853}
\author{Minwoo Jung}
\address{School of Applied and Engineering Physics, Cornell University, Ithaca, NY 14853}
\author{Gennady Shvets}
\address{School of Applied and Engineering Physics, Cornell University, Ithaca, NY 14853}
\author{A. Nick Vamivakas}
\address{Institute of Optics, University of Rochester, Rochester, NY, USA}
\author{Christina Manolatou}
\address{School of Electrical and Computer Engineering, Cornell University, Ithaca, NY 14853}
\email{fr37@cornell.edu}

\begin{abstract}
We present a many-body theory for exciton-trion-polaritons in doped two-dimensional semiconducting materials. Exciton-trion-polaritons are robust coherent hybrid excitations involving excitons, trions, and photons. Signatures of these polaritons have been recently seen in experiments. In these polaritons, the 2-body exciton states are coupled to the material ground state via exciton-photon interaction and the 4-body trion states are coupled to the exciton states via Coulomb interaction. The trion states are not directly optically coupled to the material ground state. The energy-momentum dispersion of these polaritons exhibit three bands. We calculate the energy band dispersions and the compositions of polaritons at different doping densities using Green's functions. The energy splittings between the polariton bands, as well as the spectral weights of the polariton bands, depend on the strength of the Coulomb coupling between the excitons and the trions and which in turn depends on the doping density.     
\end{abstract}  
                                    
\maketitle

\begin{figure}
  \begin{center}
   \includegraphics[width=1.0\columnwidth]{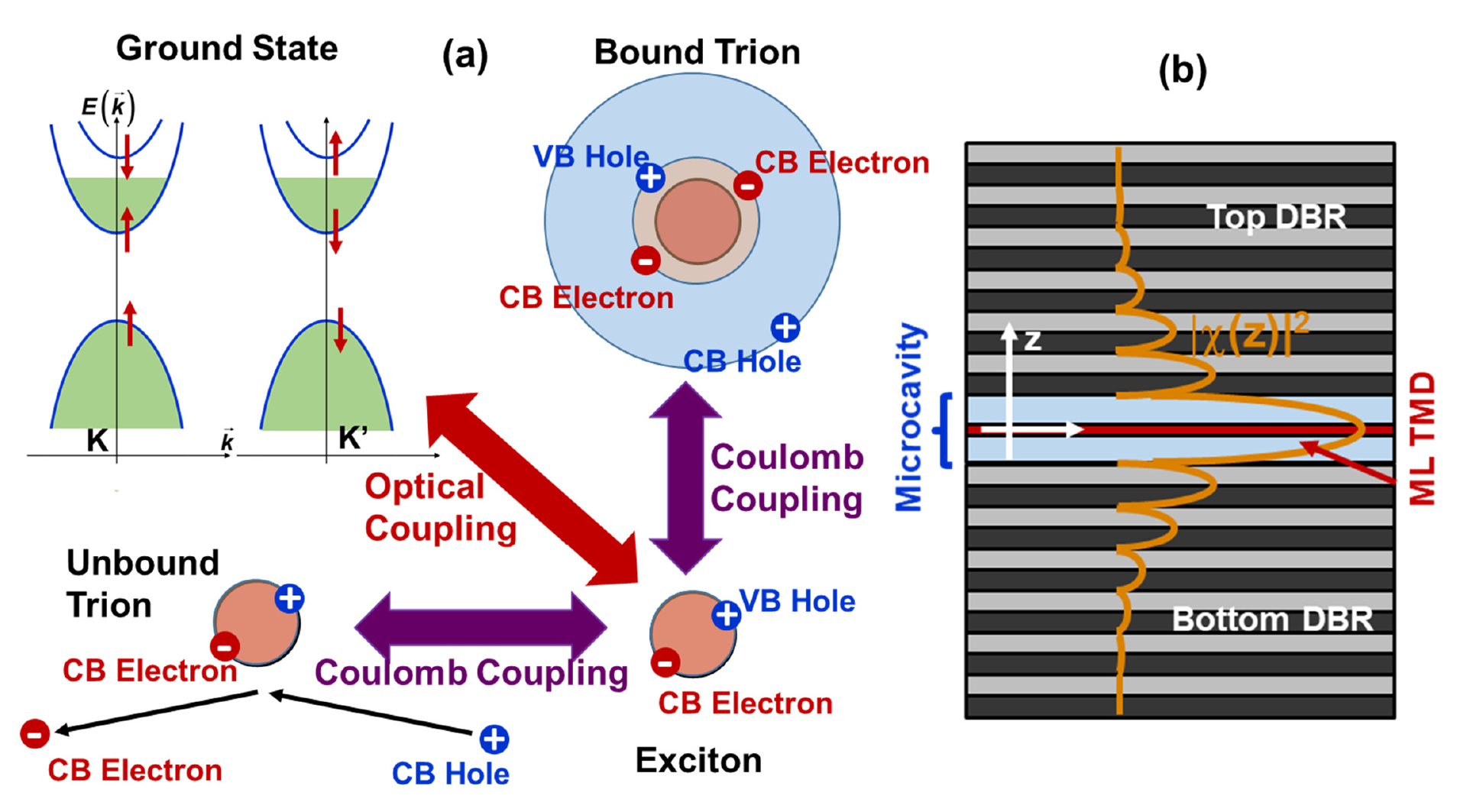}
    \caption{(a) The nature of couplings among bound and unbound trion states, exciton states, the material ground state, and photons in exciton-trion-polaritons are depicted for an electron-doped 2D material~\cite{Rana20,Rana20b}. (b) A 2D material monolayer embedded inside an optical microcavity.}
    \label{fig:fig1}
  \end{center}
\end{figure}

The scientific interest in coherent hybrid excitations of light and matter, or polaritons, stems both from a fundamental science perspective and also from practical device considerations~\cite{Pekar58,Yama1,Yama2}. Very recently, signatures of coherent hybrid excitations involving excitons, trions, and photons in doped two-dimensional (2D) materials have been reported in the literature~\cite{Imam16,Nick18,Emman20,Cuadra18,Duff17}. Although there is no consensus yet on the nature of these hybrid excitations~\cite{Imam16,Nick18,Emman20,Cuadra18,Duff17}, these experimental findings are interesting as they call into question~\cite{Imam16} the traditional description of a trion as a bound 3-body state consisting of an exciton and a free charge carrier~\cite{Combes03,Combes12,Berk13,Suris01,Urba17}. Several recent works have contributed to clarifying the nature of excitons and trions in doped semiconductors ~\cite{Rana20,Suris03,Macdonald17,Imam16,Chang19,Rana20b}. Recently, the authors have presented a model based on two coupled Schr{\"o}dinger equations to describe 2-body excitons and 4-body trions in electron-doped 2D materials~\cite{Rana20,Rana20b}. A 4-body bound trion state consists of a CB electron-hole pair bound to an exciton. The two Schr{\"o}dinger equations are coupled as a result of Coulomb interactions between the excitons and the trions in doped materials. Good approximate eigenstates of the coupled system can be constructed from superpositions of exciton and trion states. This superposition includes both bound trion states as well as unbound trion states. The latter are exciton-electron scattering states. These superposition states, first proposed by Suris~\cite{Suris03}, resemble the exciton-polaron variational states proposed by Sidler et al.~\cite{Imam16,Macdonald17}. Furthermore, the two prominent peaks observed in the optical absorption spectra of doped 2D materials do not correspond to pure exciton or pure trion states as is often assumed. Each peak corresponds to a state which is superposition of exciton and trion states~\cite{Rana20}. The model developed by the authors~\cite{Rana20,Rana20b}, rather interestingly, also showed that the 4-body trion states have no direct optical matrix elements with the material ground state. The contribution to the material optical conductivity from trion states results almost entirely from the latter's Coulomb coupling to the 2-body exciton states~\cite{Rana20b} (see Fig.~\ref{fig:fig1}(a)). Several suggested approaches~\cite{Nick18,Emman20,Cuadra18,Duff17,Kyriienko20} towards understanding exciton-trion-polaritons suffer from conceptual errors by i) assuming direct optical matrix element between the trion state and the ground state and ii) ignoring Coulomb coupling between the trion and exciton states. 

In this paper, we present a many-body theory for exciton-trion-polaritons in 2D materials based on our prior work on excitons and trions in electron-doped 2D materials~\cite{Rana20,Rana20b}. The results presented in this paper clarify the nature of exciton-trion-polaritons and are expected to stimulate further studies of these highly correlated states of light and matter. We describe the basic physics underlying these polaritons, calculate their energy dispersions, and figure out their compositions at different doping densities. Since the 4-body trion states also include the continuum of exciton-electron scattering states (or unbound trion states), the polariton problem requires a many-body approach for its complete and accurate description. We show here that the optical coupling between the excitons and the material ground state and the Coulomb coupling between the trions and the excitons result in robust exciton-trion-polaritons that exhibit three bands in their energy-momentum dispersion. The energy splittings between these polariton bands, as well as the spectral weights of these bands, depend on the strength of the Coulomb coupling between the excitons and the trions and which in turn depends on the doping density. Furthermore, exciton-electron scattering, which is inevitable at large electron densities, results in large broadening of the polariton band closest in enery to the continuum of exciton-electron scattering states (or unboud trion states).    

 Although the focus in this paper will be on electron-doped 2D transition metal dichalcogenides (TMD) materials, the arguments are kept general enough to be applicable to any 2D material. We consider a 2D material monolayer embedded inside an optical microcavity (Fig.\ref{fig:fig1}(b)). The relevant cavity optical modes is assumed to be transverse (no polarization component out of the plane of the 2D material). The Hamiltonian describing electrons and holes in the TMD layer (near the $K$ and $K'$ points in the Brillouin zone) interacting with each other and with an optical mode of in-plane momentum $\vec{Q}$ in the rotating wave approximation is~\cite{Xiao12,Changjian14,HWang16,Mano16,Rana20,Rana20b},
\begin{eqnarray}
H & = & \sum_{\vec{k},s} E_{c,s}(\vec{k}) c_{s}^{\dagger}(\vec{k})c_{s}(\vec{k}) + \sum_{\vec{k},s} E_{v,s}(\vec{k}) b_{s}^{\dagger}(\vec{k})b_{s}(\vec{k}) \nonumber \\
& + & \frac{1}{A}\sum_{\vec{q},\vec{k},\vec{k}',s,s'} U(q)  c_{s}^{\dagger}(\vec{k}+\vec{q})b_{s'}^{\dagger}(\vec{k}'-\vec{q})b_{s'}(\vec{k}')c_{s}(\vec{k}) \nonumber \\
& + & \frac{1}{2A}\sum_{\vec{q},\vec{k},\vec{k}',s,s'} V(q)  c_{s}^{\dagger}(\vec{k}+\vec{q})c_{s'}^{\dagger}(\vec{k}'-\vec{q})c_{s'}(\vec{k}')c_{s}(\vec{k}) \nonumber \\
& + & \hbar \omega(\vec{Q}) a^{\dagger}(\vec{Q})a(\vec{Q}) \nonumber \\
& + & \frac{1}{\sqrt{A}}\sum_{\vec{k,s}} \left( g_{s}c_{s}^{\dagger}(\vec{k}+\vec{Q})b_{s}(\vec{k})a(\vec{Q})  + h.c \right)
\label{eq:H}
\end{eqnarray}
Here, $E_{c,s}(\vec{k})$ and $E_{v,s}(\vec{k})$ are the conduction and valence band energies. $s,s'$ represent the spin/valley  degrees of freedom in the 2D material. $s=\{\sigma,\tau\}$, where $\sigma = \pm 1$ and $\tau = \pm 1$ represent spin and valley degree of freedom, respectively. We assume for simplicity that the electron and hole effective masses, $m_{e}$ and $m_{h}$, respectively, are independent of $s$. $U(\vec{q})$  represents Coulomb interaction between electrons in the CB and VB and $V(\vec{q})$ represents Coulomb interaction among the electrons in the CB. $\hbar \omega(\vec{Q})$ is the energy of a photon with in-plane momentum $\vec{Q}$, and $g_{s}$ is the electron-photon coupling constant. $g_{s}$ is assumed to be non-zero only for the case of the optical coupling between the top most valence band and the conduction band of the same spin (for $s=\{+1,+1\}$ or $s=\{-1,-1\}$). Other than for phase factors that are not relevant to the discussion in this paper, the non-zero values of $g_{s}$ can be written as~\cite{HWang16,Mano16}, $g = |g_{s}| = ev\chi(z=0)\sqrt{\hbar/(2\langle \epsilon \rangle\omega(\vec{Q}))}$, where, $v$ is the interband velocity matrix element~\cite{Xiao12,Changjian14,HWang16,Mano16}, $\chi (z)$ describes the amplitude of the optical mode in the z-direction (Fig.~\ref{fig:fig1}(b)), and $\langle \epsilon \rangle$ is the average dielectric constant experienced by the cavity optical mode.

The energy dispersion and the spectral weight of the exciton-trion-polaritons can be found from the poles of the retarded photon Green's function $G^{ph}(\vec{Q},t) = -(i/\hbar)\theta(t)\langle [a(\vec{Q},t), a^{\dagger}(\vec{Q},0)] \rangle$. The equation for the Green's function is,
\begin{eqnarray}
  & & \left[ \hbar \omega(\vec{Q}) - i\gamma_{p} + i\hbar \frac{\partial}{\partial t} \right] G^{ph}(\vec{Q},t) = \delta(t) \nonumber \\
  & & - \frac{\sqrt{2}g}{\sqrt{A}} \sum_{\vec{k}} G^{ex-ph}_{\vec{Q},T}(\vec{k};t)  
\end{eqnarray}
Here, $2\gamma_{p}$ is the inverse photon lifetime in the optical cavity, and, 
\begin{equation}
G^{ex-ph}_{\vec{Q},T}(\vec{k};t) =  -\frac{i}{\hbar} \theta(t) \langle \left[P^{\dagger}_{\vec{Q},T}(\vec{k};t),a^{\dagger}(\vec{Q},0) \right] \rangle 
\end{equation}
$P_{\vec{Q},T}(\vec{k};t)$ is the transverse polarization operator. In 2D TMDs, the in-plane polarized optical mode couples to excitons from both $K$ and $K'$ valleys. It is appropriate to consider superpositions of exciton states from both valleys that couple selectively to optical modes with TM or TE polarizations. This superposition state is either the longitudinal exciton (which couples only to a TM-polarized optical mode) or the transverse exciton (which couples only to a TE or in-plane polarized optical mode)~\cite{Mano16,HWang16}. For transverse excitons, the polarization operator $P_{\vec{Q},T}(\vec{k};t)$ equals, 
\begin{equation}
P_{\vec{Q},T}(\vec{k};t) = \frac{1}{\sqrt{2}}\sum_{s} \frac{g_{s}}{g} c_{s}^{\dagger}(\vec{k}+\vec{Q},t)b_{s}(\vec{k},t) 
\end{equation}
The polarization operator can be obtained from the coupled exciton and trion equations given by Rana et al.\cite{Rana20,Rana20b}. Assuming, for simplicity, that the optical mode is coupled to only the $n$-th exciton state in each valley (typically $n=0$ state, the lowest energy exciton state, is of interest), the result for the photon Green's function is found to be,
\begin{equation}
[G^{ph}(\vec{Q},\omega)]^{-1} = \hbar \omega - \hbar \omega(\vec{Q}) + i\gamma_{p} - \Sigma^{ph}(\vec{Q},\omega) 
\end{equation}
where photon self-energy $\Sigma^{ph}(\vec{Q},\omega)$ is,
\begin{eqnarray}
  & &   \Sigma^{ph}(\vec{Q},\omega) = \sum_{s} \slashed{G}^{ex}_{n,s}(\vec{Q},\omega)   \nonumber \\
  & & \times \left|  g_{s} \int \frac{\displaystyle d^{2}\vec{k}}{\displaystyle (2\pi)^{2}} \phi^{ex}_{n,\vec{Q}}(\vec{k}+ \lambda_{h}\vec{Q}) \sqrt{1 - f_{c,s}(\vec{k}+\vec{Q})} \right|^{2}
   \label{eq:selfph}
\end{eqnarray}
Here, $\phi^{ex}_{n,\vec{Q}}(\vec{k}+ \lambda_{h}\vec{Q})$ is the eigenfunction of the $n$-th exciton state~\cite{Rana20,Rana20b}. $\lambda_{h} = 1-\lambda_{e} = m_{h}/m_{ex}$, $m_{ex} = m_{e} + m_{h}$, and $f_{c,s}(\vec{k})$ is the occupation probability for the CB electron states. The bare exciton Green's function $\slashed{G}^{ex}_{n,s}(\vec{Q},\omega)$ (which does not include contribution to the exciton self-energy from exciton-photon interaction) appearing in (\ref{eq:selfph}) is,
\begin{equation}
[\slashed{G}^{ex}_{n,s}(\vec{Q},\omega)]^{-1} = \hbar \omega - E^{ex}_{n,s}(\vec{Q}) + i\gamma_{ex} - \left. \Sigma^{ex}_{n,s}(\vec{Q},\omega) \right|_{tr}  
\end{equation}
In the above expression, $E^{ex}_{n,s}(\vec{Q})$ is the energy of the $n$-th exciton state of spin/valley $s$~\cite{Rana20,Rana20b}, $\gamma_{ex}$ describes the rate of coherence decay of the exciton polarization due to all processes other than exciton-electron scattering. The latter is included explicitly in the exciton self-energy $\Sigma^{ex}_{n,s}(\vec{Q},\omega)|_{tr}$~\cite{Rana20,Rana20b} resulting from exciton-electron interaction. Exciton-electron interaction can be described in terms of exciton-trion coupling~\cite{Rana20,Rana20b}, including couplings to both bound and unbound 4-body trion states. The latter are just exciton-electron scattering states (Fig.~\ref{fig:fig1}(a)). Expression for the exciton self-energy was found by Rana et al.~\cite{Rana20},
\begin{eqnarray}
 \left. \Sigma^{ex}_{n,s}(\vec{Q},\omega) \right|_{tr} & = & \sum_{m,s'} \left. \Sigma^{ex}_{n,m,s,s'}(\vec{Q},\omega) \right|_{tr} \nonumber \\
&  = & \sum_{m,s'} \frac{ (1 + \delta_{s,s'}) \, \left| M_{n,m,s,s'}(\vec{Q}) \right|^{2}}{\hbar\omega - E^{tr}_{n,m,s,s'}(\vec{Q}) + i\gamma_{tr}} \label{eq:selfex}
\end{eqnarray}
The expressions for the Coulomb matrix elements $M_{n,m,s,s'}(\vec{Q})$, coupling 2-body exciton states with spin/valley $s$ to 4-body trion states with spin/valley $s,s'$, can be found in a previous paper by Rana et al.~\cite{Rana20}. The summation over $m$ above implies a summation over all bound and unbound 4-body trion states consistent with the values of $s$ and $s'$. $E^{tr}_{n,m,s,s'}(\vec{Q})$ is the energy of a 4-body trion state and $\gamma_{tr}$ is a phenomenological parameter describing the decay of the coherence of four-body correlations. $\Sigma^{ex}_{n,s}(\vec{Q},\omega) |_{tr}$ is roughly proportional to the doping density~\cite{Rana20}. Not surprisingly, the photon self-energy in (\ref{eq:selfph}) can be written in terms of the optical conductivity of the 2D material~\cite{Rana20,Rana20b},
\begin{equation}
  \Sigma^{ph}(\vec{Q},\omega) = -i\hbar \frac{|\chi(z=0)|^{2}}{2 \langle  \epsilon \rangle} \sigma(\vec{Q},\omega)
  \end{equation}
Although the dispersion of the exciton-trion-polaritons can be obtained from the poles of the photon Green's function, Hopfield coefficients~\cite{Hop58,haugbook} play an important role in describing the composition of polariton states. The same information is  also provided by the spectral density functions, which we discuss next. The photon spectral density function $S^{ph}(\vec{Q},\omega)$ equals $-2 \hbar \text{Im}\left\{ G^{ph}(\vec{Q},\omega) \right\}$. The spectral density $S^{ex}_{n,T}(\vec{Q},\omega)$ of the transverse exciton equals $-2 \hbar \text{Im}\left\{ G^{ex}_{n,T}(\vec{Q},\omega) \right\}$. Assuming $E^{ex}_{n,s}(\vec{Q})=E^{ex}_{n,-s}(\vec{Q})$ and $|g_{s}|=|g_{-s}|$, the transverse exciton Green's function $G^{ex}_{n,T}(\vec{Q},\omega)$ is found to be,
\begin{eqnarray}
  [G^{ex}_{n,T}(\vec{Q},\omega)]^{-1} & = & \hbar \omega - E^{ex}_{n,s}(\vec{Q}) + i\gamma_{ex} - \left. \Sigma^{ex}_{n,s}(\vec{Q},\omega) \right|_{tr} \nonumber \\
  & & - \left. \Sigma^{ex}_{n,T}(\vec{Q},\omega) \right|_{ph} 
\end{eqnarray}
The spin/valley index $s$ on the right hand side stands for any one of the two values for which $|g_{s}|\ne 0$, and the exciton-photon interaction contribution to the transverse exciton self-energy is,  
\begin{eqnarray}
  & & \Sigma^{ex}_{n,T}(\vec{Q},\omega)|_{ph} = \nonumber \\
  & & \sum_{s} \frac{ \left|  g_{s} \int \frac{\displaystyle d^{2}\vec{k}}{\displaystyle (2\pi)^{2}} \phi^{ex}_{n,\vec{Q}}(\vec{k}+ \lambda_{h}\vec{Q}) \sqrt{1 - f_{c,s}(\vec{k}+\vec{Q})} \right|^{2}   }{\hbar \omega - \hbar \omega(\vec{Q}) + i\gamma_{p} }  \nonumber \\
\end{eqnarray}
We assume that only a single bound 4-body singlet trion state of index $m$ exists ($m=0$ implies the lowest energy bound trion state), and it exists only when the exciton and the bound CB electron-hole pair pair belong to different valleys and have different spins~\cite{Rana20}. We define a 4-body bound transverse trion state as the one formed by the binding of a CB electron-hole pair to a transverse exciton~\cite{Rana20}. Finally, the spectral density function for the bound transverse trion state is $S^{tr}_{n,m,T}(\vec{Q},\omega) = -2 \hbar \text{Im}\left\{ G^{tr}_{n,m,T}(\vec{Q},\omega) \right\}$, where the Green's function of the 4-body bound transverse trion state is,
\begin{eqnarray}
  [G^{tr}_{n,m,T}(\vec{Q},\omega)]^{-1} & = & \hbar \omega - E^{tr}_{n,m,s,-s}(\vec{Q}) + i\gamma_{tr} \nonumber \\
  & & - \Sigma^{tr}_{n,m,T}(\vec{Q},\omega)
  \label{eq:tr}
\end{eqnarray}
where,
\begin{eqnarray}
  && \Sigma^{tr}_{n,m,T}(\vec{Q},\omega) = \nonumber \\
  & & \frac{  \, \left| M_{n,m,s,-s}(\vec{Q}) \right|^{2}}{\splitfrac{\hbar\omega - E^{ex}_{n,s}(\vec{Q}) + i\gamma_{ex} - \Sigma^{ex}_{n,T}(\vec{Q},\omega)|_{ph}}{ - {\displaystyle \sum_{m'\ne m,s'} } \Sigma^{ex}_{n,m',s,s'}(\vec{Q},\omega)|_{tr}}} \nonumber \\
  \label{eq:selftr}
\end{eqnarray}
As before, the spin/valley index $s$ on the right hand sides in (\ref{eq:tr}) and (\ref{eq:selftr}) stands for any one of the two values for which $|g_{s}|\ne 0$. 
\begin{figure}
  \begin{center}
   \includegraphics[width=0.87\columnwidth]{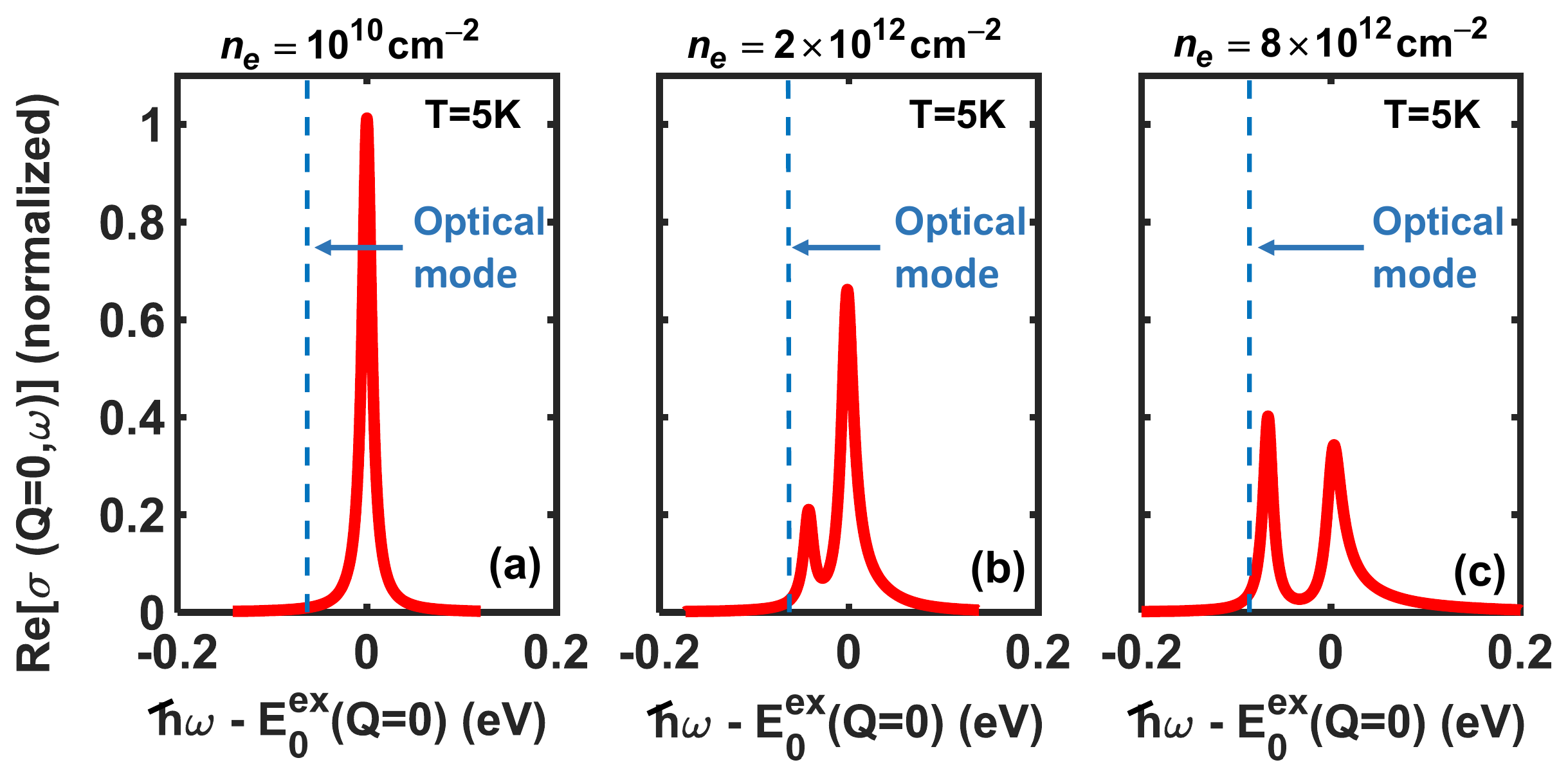}
   \caption{Calculated real part of the optical conductivity, $\sigma(\vec{Q}=0,\omega)$, for in-plane (TE) light polarization, is plotted for three different electron densities ($n_{e} = 10^{10}, 2\times 10^{12},8\times 10^{12}$ cm$^{-2}$) for electron-doped monolayer 2D MoSe$_{2}$. Only the lowest energy exciton state is considered in the calculations. The spectra are all normalized to the peak optical conductivity value at zero electron density. T = 5K. The frequency axis is offset by the exciton energy $E^{ex}_{n=0,s}(\vec{Q}=0)$. The position of the cavity optical mode is also indicated (see Fig.~\ref{fig:fig4}). Two prominent peaks are seen in the absorption spectra when the electron density exceeds $\sim 10^{12}$ cm$^{-2}$. Each peak corresponds to a state that is a superposition of exciton and trion states~\cite{Rana20}. The spectral weight shifts from  the higher energy peak to the lower energy peak with the increase in the electron density.}  
    \label{fig:fig3}
  \end{center}  
\end{figure}
\begin{figure}[h]
  \begin{center}
   \includegraphics[width=0.85\columnwidth]{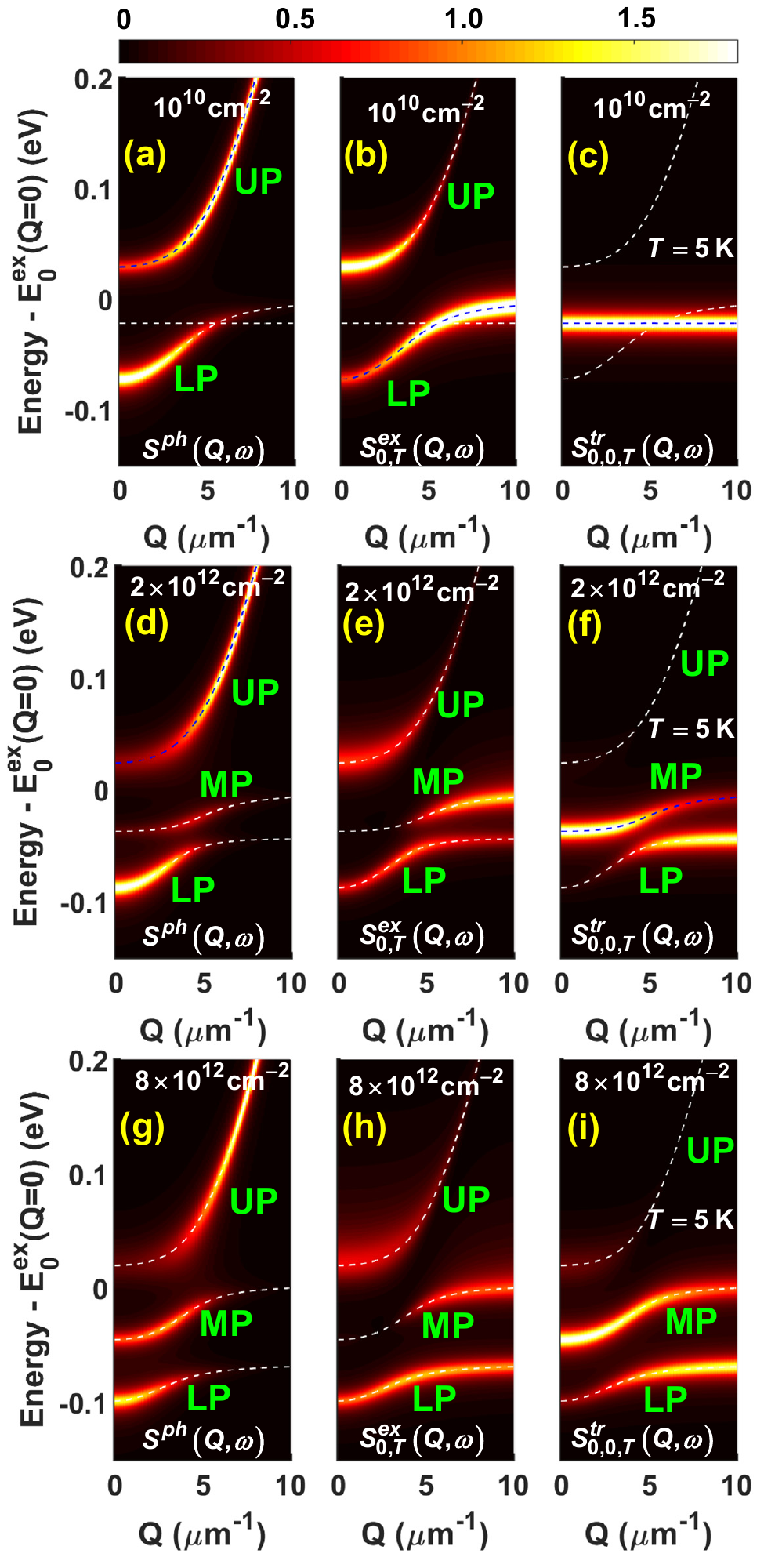}
   \caption{Calculated exciton-trion-polariton energy dispersions (dashed lines) and the spectral densities of the photon ($S^{ph}(\vec{Q},\omega)$), the transverse exciton ($S^{ex}_{n=0,T}(\vec{Q},\omega)$), and the transverse bound trion ($S^{tr}_{n=0,m=0,T}(\vec{Q},\omega)$), are plotted for three different electron densities ($n_{e} =10^{10}, 2\times 10^{12},8\times 10^{12}$ cm$^{-2}$) for an electron-doped monolayer 2D MoSe$_{2}$ inside an optical cavity (Fig.\ref{fig:fig1}(b)). In each case, the cavity optical mode is tuned $\sim 20$ meV below the lower energy peak in the optical absorption spectra (as indicated in Fig.~\ref{fig:fig3}). T=5K. The unit in the colorbar is $10^{-13}$ s.}
    \label{fig:fig4}
  \end{center}
\end{figure}

For simulations, we consider an electron-doped monolayer of 2D MoSe$_{2}$ inside an optical microcavity, as shown in Fig.~\ref{fig:fig1}(b). In monolayer MoSe$_{2}$, spin-splitting of the conduction bands is large ($\sim$35 meV~\cite{Kosmider13}) and the lowest conduction band in each of the $K$ and $K'$ valleys is optically coupled to the topmost valence band~\cite{Xiao13}. We use effective mass values of $0.7 m_{o}$ for both $m_{e}$ and $m_{h}$ which agree with the recently measured value of $0.35 m_{o}$ for the exciton reduced mass~\cite{Goryca19}. The in-plane polarized (TE) cavity optical mode has a parabolic dispersion and corresponds to a photon mass of $10^{-5} m_{o}$. $|\chi (z=0)|^{2} = 10$ $\mu$m$^{-1}$. We use a wavevector-dependent dielectric constant $\epsilon(\vec{q})$, appropriate for 2D materials~\cite{Changjian14,Rana20}, to screen the Coulomb potentials. We assume that $\gamma_{ex}=\gamma_{tr}=\gamma_{p} \sim 6$ meV~\cite{Knorr16}. We compute exciton and trion eigenfunctions and eigenenergies for different momenta and electron densities as described by Rana et al.~\cite{Rana20}.

Fig.~\ref{fig:fig3} shows the real part of the optical conductivity (optical absorption spectra) for three different electron densities and Fig.~\ref{fig:fig4} shows the corresponding polariton dispersions (dashed lines) as well as the spectral densities of the photon, the transverse exciton, and the transverse bound trion. We assume in simulations that the cavity optical mode is tuned $\sim 20$ meV below the lower energy peak in the optical absorption spectra (as indicated in Fig.~\ref{fig:fig3}. At the lowest electron density ($n=10^{10}$ cm$^{-2}$), the lower energy peak in the optical absorption spectrum has essentially no optical oscillator strength and all the spectral weight lies in the higher energy peak (which is the only one seen in Fig.~\ref{fig:fig3}(a)). The higher and lower energy states at such small electron densities correspond to essentially pure exciton and pure (bound) trion states, respectively~\cite{Rana20}. The resulting polariton dispersion, not surprisingly, shows two bands, UP (upper polariton) and LP (lower polariton), which represent exciton-polaritons (Fig.\ref{fig:fig4}(a,b)). The bound trion states do not form polaritons as they have no oscillator strength. When the electron density increases beyond $\sim 10^{12}$ cm$^{-2}$, exciton and trion states become coupled as a result of strong Coulomb interactions, and the resulting optical absorption spectra show two prominent peaks (Fig.~\ref{fig:fig3}(b)). Each peak corresponds to a state that is a superposition of 2-body exciton and 4-body (bound) trion states~\cite{Rana20}. The polariton dispersion for $n = 2\times 10^{12}$ cm$^{-2}$ shows three bands, UP, MP (middle polariton), and LP (Fig.\ref{fig:fig4}(d,e,f)). The Rabi splitting between the LP and MP bands is however small and reflects the fact that the lower energy peak in the optical absorption spectra (Fig.~\ref{fig:fig3}(b)) does not have much optical oscillator strength. As the electron density increases further, the spectral weight continues to shift from the higher energy peak in the absorption spectrum to the lower energy peak and, in addition, the higher energy peak broadens, becomes non-Lorentzian, and develops a pedestal as a result of exciton-electron scattering  (i.e., Coulomb coupling of the exciton and unbound trion states). This pedestal is visible on the higher energy side of the peak in Fig.~\ref{fig:fig3}(c) for $n = 8\times 10^{12}$ cm$^{-2}$. When $n = 8\times 10^{12}$ cm$^{-2}$, the increase in the oscillator strength of the lower energy peak is reflected in the large Rabi splitting between the LP and MP polariton bands in Fig.~\ref{fig:fig4}(g,h,i). Also visible in Fig.~\ref{fig:fig4}(g,h,i) is the extremely large broadening of the UP band from dephasing caused by exciton-electron scattering (or coupling between exciton and unbound trions) at this large doping density. The spectral densities obey the following sum rule,
\begin{equation}
  \int \frac{d\omega}{2\pi} \left[ S^{ph}(\vec{Q},\omega) + S^{ex}_{n=0,T}(\vec{Q},\omega) + S^{tr}_{n=0,m=0,T}(\vec{Q},\omega) \right] = 1
\end{equation}
where the frequency integral is restricted to any one of the three polariton bands.

To the best of our knowledge, only one experimental work by Dhara et al. has reported exciton-trion-polariton energy-momentum dispersion~\cite{Nick18}. Dhara et al. reported a negative mass dispersion for the LP band which does not agree with theoretical model presented here. According to the model presented in this paper, the extremely small positive mass of the cavity optical mode will result in a positive mass energy-dispersion for all three polariton bands irrespective of the energy-momentum dispersions of excitons and trions. We expect that the work presented in this paper wil stimulate further exploration of exciton-trion-polaritons in 2D materials.        

The authors would like to acknowledge helpful discussions with Francesco Monticone, and support from CCMR under NSF-NRSEC grant number DMR-1719875 and NSF EFRI-NewLaw under grant number 1741694.

\end{document}